\documentclass[ammsymb,
amssymb,
twocolumn,
prb,
superscriptaddress,
floats,
showkeys,
showpacs,
a4paper]{revtex4-1}
  
\usepackage[T1]{fontenc}\usepackage{bm}
\usepackage{graphicx}
\usepackage{natbib}
\usepackage{epstopdf}
\usepackage[euler]{textgreek}
\usepackage{ulem}
\usepackage{float}
\usepackage{color}

\begin{document}
 
\title{Quadexciton cascade and fine structure splitting of the triexciton in a single quantum dot}

\author{M. R. Molas}
\email{maciej.molas@gmail.com}
\affiliation{Faculty of Physics, University of Warsaw, ul. Pasteura 5, 02-093 Warszawa, Poland}
\affiliation{Laboratoire National des Champs Magn\'etiques Intenses, CNRS-UJF-UPS-INSA, 25, avenue des Martyrs, 38042 Grenoble, France} 
\author{A. A. L. Nicolet}
\affiliation{Laboratoire National des Champs Magn\'etiques Intenses, CNRS-UJF-UPS-INSA, 25, avenue des Martyrs, 38042 Grenoble, France} 
\author{A. Babi\'nski}
\affiliation{Faculty of Physics, University of Warsaw, ul. Pasteura 5, 02-093 Warszawa, Poland}
\author{M. Potemski}
\affiliation{Laboratoire National des Champs Magn\'etiques Intenses, CNRS-UJF-UPS-INSA, 25, avenue des Martyrs, 38042 Grenoble, France} 

\date{\today}

\begin{abstract}

We report the properties of emission lines associated with the cascaded recombination of a quadexciton in single GaAlAs/AlAs quantum dots, studied by means of polarization-resolved photoluminescence and single-photon correlation experiments. It is found that photons which are emitted in a double-step 4X-3X process preserve their linear polarization, similarly to the case of conserved polarization of correlated photons in the 2X-X cascade. In contrast, an emission of either co-linear or cross-linear pairs of photons is observed for the 3X-2X cascade. Each emission line associated with the quadexciton cascade shows doublet structure in the polarization-resolved photoluminescence experiment. The maximum splitting is seen when the polarization axis is chosen along and perpendicular to the [110] crystallographic direction. This effect is ascribed to the fine structure splitting of the exciton and triexciton states in the presence of an anisotropic confining potential of ae dot. We also show that the splitting in the triexciton state surpasses that in the exciton state by a factor up to eight and their ratio scales with the energy distance between the 3X and X emission lines, thus, very likely, with a lateral size and/or a composition of the dot.

\end{abstract}

\maketitle

\section{Introduction \label{sec:Intro}}

Semiconductor quantum dots (QDs), when viewed as representative single-photon emitters, display a unique property of the cascaded (one after another) photon emission. Best known example of such a cascade is a consecutive recombination of a single electron-hole (\mbox{$e$-$h$}) pair from the biexciton state to the exciton state, the emission process denoted as 2X, followed by the radiative decay of the latter to the ground state (empty dot), labelled X. The 2X-X photon cascade might be a source of entangled photons pairs\cite{benson,stevenson,akopian} what invokes possible applications in quantum-cryptography\cite{gisin} and/or -teleportation.\cite{jennevein} A pertinent obstacle which appears in this context is an inherent anisotropy of most of QDs, which results in the so-called fine structure splitting (FSS) of the intermediate exciton state, observed in (linear) polarization-resolved experiments.\cite{gammonprl,favero,flissikowski,bayerprb,karlsson} This splitting, overall expected for spin-unpaired states, is due to anisotropic $e$-$h$ exchange interaction. It is apparent in QDs because of the asymmetry of confining potentials, anisotropy of electronic bands and/or strain distribution. Little explored so far are cascaded emissions beyond a simple two-step (the 2X and X emission) process,\cite{karlsson,bester,singh,seguin,kadantsew} such as, a three-step recombination decay of the triexciton state or even a four-step decay of the quadexciton state.\cite{arashidacharged,perrson,arashida} The triexciton forming the spin-unpaired state which appears in these latter cascades can also be expected to exhibit a sensitivity to the anisotropic $e$-$h$ exchange interaction, similarly like the exciton state does. The properties of polarized photons emitted within such high-order cascades may bring new information on optical processes and sources of anisotropy in semiconductor QDs.

In this paper, we report on optical studies of a series of single GaAlAs/AlAs QDs. The emission lines associated with the quadexciton cascade (subsequent recombination of the quadexciton, triexciton, biexciton, and exciton state) were identified with photon correlation experiments. Photons which cascade in a double-step 4X-3X process are found to preserve their linear polarization, similarly to the case of conserved polarization of correlated photons in the 2X-X cascade. The linear polarization of the cascaded photons is however lost at the biexciton intermediate state since the consecutive emissions of both co-linear and cross-linear pairs of photons is observed in the 3X-2X cascade. Each emission line associated with the quadexciton cascade shows a doublet structure in polarization-resolved experiments. A pair of 4X and 3X emission doublets is found to follow the same polarization rules as the pair of 2X and X emission doublets. The amplitude of the doublet splitting is the same within each pair what points out the absence of the FSS not only in the biexciton but in the quadexciton state, as well. The FSS in the triexciton state surpasses the FSS in the exciton state by a factor up to eight and their ratio scales with the energy distance between the 3X and X emission lines, thus, very likely, with a lateral size and/or a composition of the dot.

\section{Experimental details \label{sec:procedure}}

The active part of the structure used in our study was intentionally designed as a type-II GaAs/AlAs bilayer ($d_{\textrm {GaAs}}$ = 2.4 nm, $d_{\textrm {AlAs}}$ = 10 nm) embedded  between wide (100 nm) $\textrm{Ga}_{0.67}\textrm{Al}_{0.33}\textrm{As}$ barriers.\cite{truby,wysmolekappa} Previous research\cite{wysmolekE,wysmolekappa} showed that the bilayer is not perfect in the lateral directions: the Ga-rich inclusions, which can be seen as islands of $\textrm{Ga}_{1-x}\textrm{Al}_{x}\textrm{As}$  ($x$ < 0.33) replacing the original GaAs/AlAs bilayer, exist in this structure and possess all attributes of relatively strongly confined semiconductor QDs. These dots show a remarkably low surface density, at the level of $10^5-10^6$ $\textrm{cm}^{-2}$. Their emission spectra are dispersed over a wide energy range, 1.56-1.68 eV,\cite{wysmolekappa,molasnatural,martin,molascharged,pietkaprb} due to the spread in the lateral extent of the confining potential and different Al-composition.

Single dot measurements were carried out at liquid helium temperature using a typical setup for micro-photoluminescence (micro-PL) experiments. The setup includes a cold finger cryostat with short-distance optical access, a microscope objective to focus the excitation and collect the emission within the 1 \textmu m$^2$ spot, 0.5 m-long monochromators equipped with the charge-couple-device cameras (CCDs), and avalanche photodiodes (APDs). A tunable Ti:Sapphire laser was set at $\lambda$= 725 nm to assure the quasi-resonant excitation conditions, $i.e.$, to inject the electron-hole ($e$-$h$) pairs directly into QDs.\cite{kazimierczuknon} Polarisation-resolved emission was measured using a motorised, rotating half-wave plate combined with a fixed linear polariser, placed in front of the spectrometers.\cite{molasjap} For the purpose of photon correlation experiments, the micro-PL was collected via the same microscope objective but afterwards separated into two beams by a 50/50 beam splitter. Each beam was sent through a monochromator; the emission lines of interest were initially identified using the CCD cameras and then the detection was switched to APDs. APDs were triggering a start-stop measurement in the Hanbury-Brown and Twiss configuration,\cite{twist,sufczynski} and were coupled to a time-correlated coincidence counter to create the histogram of the measured time separation $\tau$. Such a histogram represents a second-order correlation function $\textrm{g}^{(2)}(\tau)$, which reflects the temporal relations between the events of individual photon emissions from a single QD.

\section{Results and discussion \label{sec:pl}}
\subsection{Photoluminescence of a single QD \label{subsec:pl}}

\begin{center}
\begin{figure*}
\includegraphics[width=151mm]{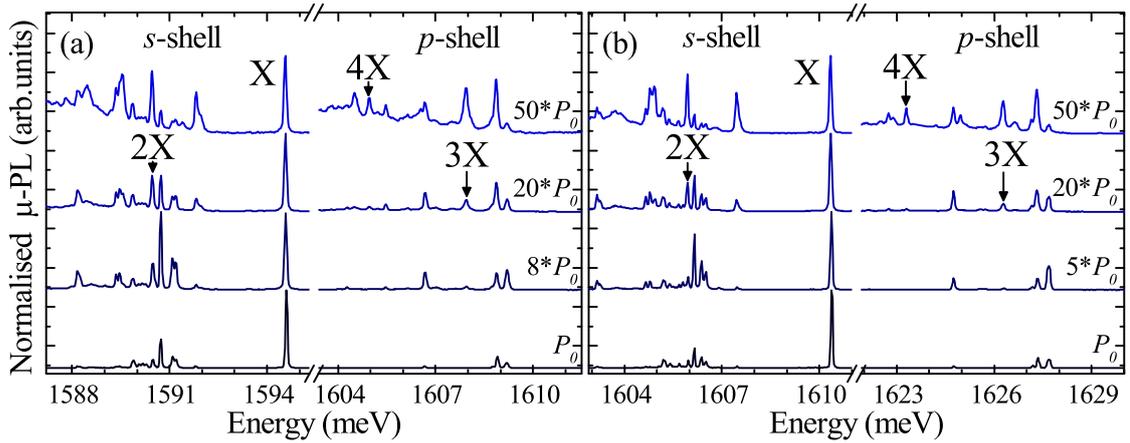}
\caption{(color online) The excitation-power evolution of the micro-PL spectra of two single GaAlAs/AlAs QDs measured at $T$=~4.2~K. The~spectra are excited quasi-resonantly (E$_{exc}$=1.71 eV). P$_0$=10 \textmu W and indicates the excitation power. The~spectra are normalized to the intensity of the X emission line and vertically shifted for clarity purpose. \label{fig:mikropl}}
\end{figure*}
\end{center}

The dots present in the investigated structure are not homogeneous and very likely characterized by different shapes/size and different chemical composition (Al content). Nevertheless their optical response is overall very similar. This is shown in Fig. \ref{fig:mikropl}(a) and (b) with the characteristic evolutions of micro-PL spectra measured as a function of the excitation power, for two selected  dots. The energy of the emission lines as well as their relative positions can be different for different dots but their spectral pattern and the sequence of appearance of subsequent lines upon increase of the excitation power is seemingly the same for each dot. More than 20 QDs were studied in micro-PL experiments (power dependence, polarization resolved). Most of photon correlation experiments were carried out on the dot which characteristic spectra are illustrated in Fig. \ref{fig:mikropl}(b), and on which we focus in the following. The emission line, labelled as X in Fig. \ref{fig:mikropl}, dominates the micro-PL spectra at low excitation powers (< 10 \textmu W). With the increase in the excitation power, new lines appear on its low energy side, and progressively, at distinctly higher energies - in the range of 13-15 meV above the X line. A rich and often complex evolution of single QD spectra, such as displayed in Fig. \ref{fig:mikropl}, results from gradual "filling" of the dot with more and more photo-created carriers, predominantly with $e$-$h$ pairs but possibly also with additional single-charge (photo-captured) carriers. When dealing with this type of spectral evolution in  strongly confined dots, one often starts reasoning in terms of "filling" the single particle (SP) levels (for electrons and holes). Those are conventionally conceived as the subsequent $s$-, $p$-, $d$-,\ldots, electronic shells (in reference to a simple, atomic-like model of a QD which results from a parabolic lateral confinement of two-dimensional carriers with parabolic dispersion relations). \cite{hawrylak,bayernature,babinskiprb2006b} A single $e$-$h$ pair in a dot, bound by the confinement and the Coulomb potential is seen as an exciton. Its ground state corresponds to the electron and hole ($e_s$,$h_s$) occupying the SP $s$-shells. A fully filled $s$-shell with two electrons and two holes form a ground (singlet) state (2$e_s$,2$h_s$) of the biexciton. The triexciton ground state (2$e_s$$e_p$,2$h_s$$h_p$) implies that carrier occupy also the higher energy $p$-shells, and this is actually a general rule that applies to the ground state (2$e_s$2$e_p$,2$h_s$2$h_p$) of the quadexciton and any higher-order multiexcitons as well. Among the multitude of lines seen in Fig. \ref{fig:mikropl}, one may expect a particular series of emission lines: X, 2X, 3X, and 4X, associated with the recombination process from the exciton, biexciton, triexciton, and quadexciton state, respectively. The identification of these emission lines, that have already been marked in Fig. \ref{fig:mikropl}, is discussed in the next sections of this paper. It is instructive to note that both the X and 2X emission lines imply the annihilation of carriers from the $s$-shells. The 4X and 3X emission lines are seen at distinctly higher energies as the processes which underlie them are associated with the annihilation of carriers from the \mbox{$p$-shells}. A dense and fairly complex pattern of emission lines seen in Fig. \ref{fig:mikropl} originates from all the recombination paths that are available to excitonic complexes in the dot. It should be pointed out that both the initial and the final recombination states may not necessarily be the ground states of multiexcitons. Their excited states, the multiplicity of which increases with a number of carriers involved, can also play such a role. Furthermore, the possibility of photo-induced capture of single charge carriers usually leads to an additional family of emission lines associated with the so-called charged excitons. Positively and negatively charged excitons are indeed observed in our spectra but discussed in detail elsewhere.\cite{molasnatural,molascharged}

\subsection{Single-photon correlations\label{subsec:correlations}}

\begin{figure}[h]
\includegraphics[width=80mm] {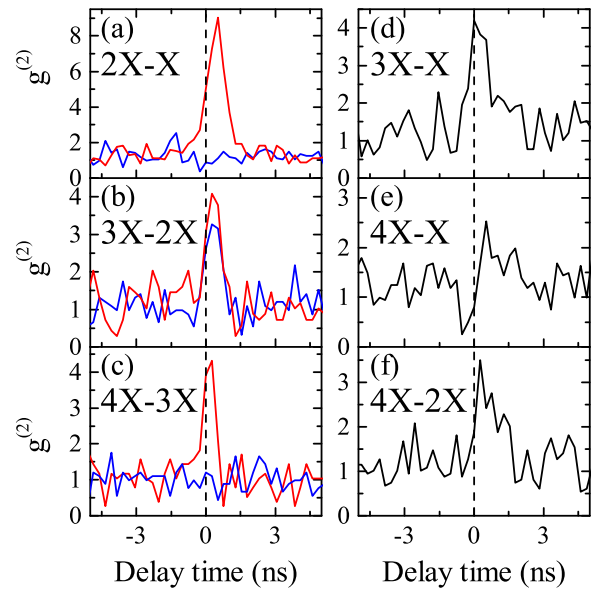}
\caption {(color online) The histograms of all the possible photon correlation measurements between the X, 2X, 3X and 4X emission lines. The red and blue curves correspond to the $\parallel$-$\parallel$ (co-polarized) and $\parallel$-$\perp$ (cross-polarized) configurations of the experiment, respectively. $\parallel$ ($\perp$) stands for the linear polarization that is parallel (perpendicular) to the [110] crystallographic direction. The black curves represent the data obtained without the polarization resolution.  The zero delay time between the start and stop beam is indicated in each panel with a black dashed line.
\label{fig:correlations}}
\end{figure}

The assignment of the series of 4X to X emission lines and the identification of the quadexciton cascade is strongly supported by photon correlation experiments. Of particular interest are cross-correlation histograms which have been investigated for the subsequent pairs of the emission lines: 4X-3X, 3X-2X, and 2X-X. In the experiments, the analysed photons were passed through linear polarisers in order to cross-correlate each of these pairs in two distinct configurations of co- and cross-polarization of the NX and (N-1)X photons (see next paragraph for the choice of the specific direction, of the polarization of light, along the [110] crystallographic axis). The resulting histograms are shown in Fig.~\ref{fig:correlations}(a)-(c). The appearance of the bunching peak at small but clearly positive delay times in the histograms for each of these pairs captures the prominent property of the sequence of the 4X-, 3X-, 2X- and X-photons being visibly emitted one after another. The extra histograms shown in Fig.~\ref{fig:correlations}(d)-(f), (no polarization resolution in these measurements) show, as expected, that 3X-photons precede X-photons, that 4X-photons precede X-photons and 4X-photons come before the 2X photons (see photon bunching peaks at positive delay times in all 3X-X, 4X-X and 4X-2X histograms).

The analysis of the polarization resolved data provides further information on the investigated four photon cascade. As shown in Fig.~\ref{fig:correlations}(a)-(c), the co- ($\parallel$-$\parallel$) and cross-polarized ($\parallel$-$\perp$) configurations produce qualitatively different histograms when correlating the 2X-X as well as the 4X-X pair. For each of these two pairs, the cascaded emission appears only in the co-polarized configuration (bunching peak at positive time delays), whereas no clear correlation or rather weakly pronounced antibunching deep can be concluded in the case of histograms investigated in the cross-polarized configuration (see Fig.~\ref{fig:correlations}(a) and (c)). Such asymmetry in the co- and cross-polarized histograms is a known property of the biexciton-exciton cascade (our 2X-X sequence) and is a consequence of the FSS splitting of the excitonic (spin unpaired) state.\cite{santori} It is logical to assume that a similar FSS splitting may also be characteristic of the triexciton, also a spin-unpaired state. Thus one may expect to observe the asymmetry in the co-and cross-polarised histograms for the quadexciton to triexciton cascade (our 4X-3X sequence) as well. However, this asymmetry should be absent in the case of the triexciton to biexciton cascade (as seen for our 3X-2X sequence - Fig.~\ref{fig:correlations}(b)) since the FSS splitting is not expected for the spin-compensated biexciton state.

\begin{figure}
\includegraphics[width=70mm]{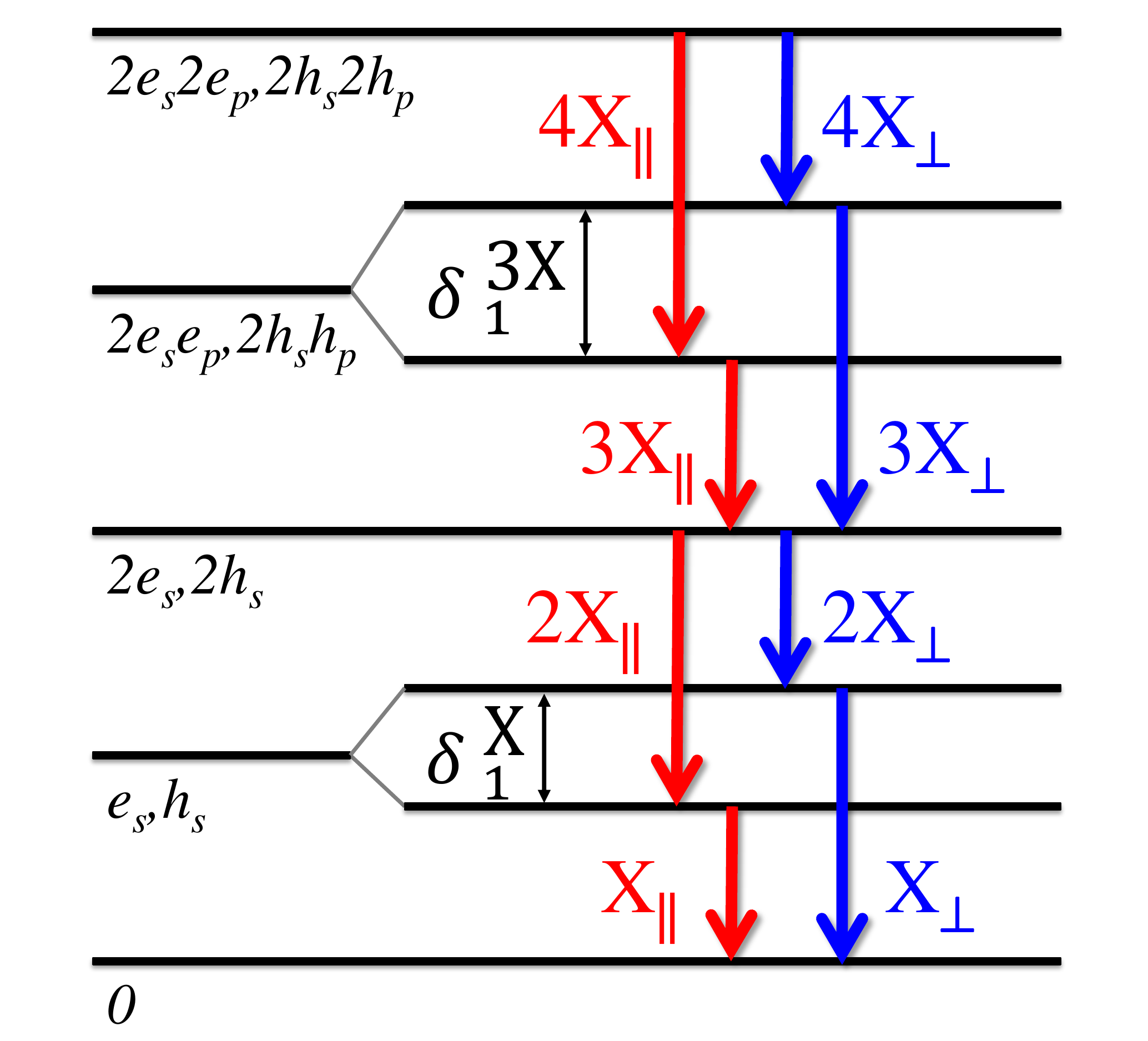}
\caption {(color online) Schematic diagram of the cascade decay of a quadexciton in a single QD. The solid blue (red) arrows denote the radiative transitions in two perpendicular polarizations.  The $\parallel$ and $\perp$ are the parallel and perpendicular linear polarizations with respect to the [110] crystallographic direction, respectively.\label{fig:scheme}}
\end{figure}

Altogether, the characteristics of photon correlation histograms shown in Fig.~\ref{fig:correlations}, confirms our assignment of the 4X, 3X, 2X, X emission lines as due to quadexciton cascade. The proposed scheme of this cascade, the multiexcitonic states and recombination processes involved, is presented in Fig.~\ref{fig:scheme}. Following the above discussion, we assume (in the first approximation) that neither quadexciton nor biexciton states are split, whereas both exciton and triexciton states exhibit the FSS splitting, in general different, correspondingly $\delta_{\textrm {1}}^{\textrm X}$ and  $\delta_{\textrm {1}}^{\textrm {3X}}$ for these two states.

\subsection{Fine structure splitting \label{subsec:fss}}

\begin{figure}
\includegraphics[width=71mm] {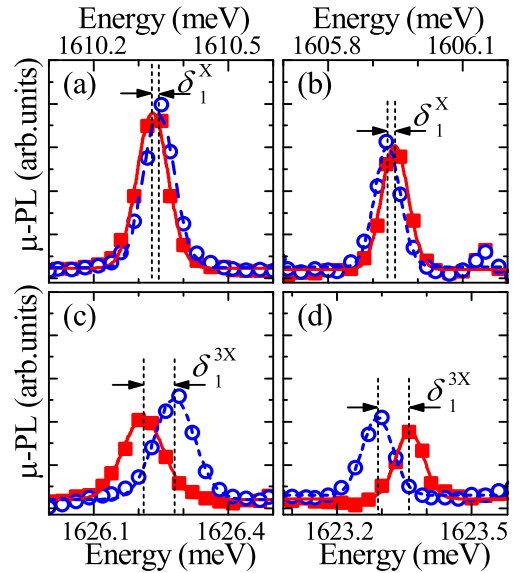}
\caption {(color online) High-resolution polarization-resolved micro-PL spectra of the emission lines attributed to the neutral exciton (a), biexciton (b), triexciton (c) and quadexciton (d) in a single GaAlAs QD measured in two polarizations: parallel (red) and perpendicular (blue) to the [110] crystallographic direction. The open circles and closed squares indicate the experimental data. The curves drawn with solid and dashed lines represent the results of fitting the data with the Gaussian function. }\label{fig:anizomikropl} 
\end{figure}

Following the proposed scheme of the quadexciton cascade (see Fig.~\ref{fig:scheme}), all four 4X, 3X, 2X and X emission lines are expected to exhibit the splitting in polarization resolved micro-PL experiments. Such experiments were performed on more than 20 different dots and each time all four components of the quadexciton cascade were probed. The characteristic data obtained for a selected dot is shown in Fig.~\ref{fig:anizomikropl}. As can be seen in this figure, each of the 4X, 3X, 2X X emission line displays two-components (splitting) which are best resolved when the emission is linearly polarized along or perpendicular to the [110] crystallographic direction of the GaAs substrate. This direction was found to be common for all investigated dots and has been concluded from the experiments in which a large ensemble of spectra was measured while progressively rotating the polarization direction of the emitted light. As seen in Fig.~\ref{fig:anizomikropl}, the amplitude of the observed splitting is of the order of 10 \textmu V for the X and 2X emission lines and somehow larger for the 4X and 3X lines. As discussed above the appearance of two distinct linearly polarized components of the investigated lines is due to fine structure splitting of the excitonic states, $i.e.$, single exciton and triexciton states in our case.

The FSS of the exciton in a QD is a result of the $e$-$h$ exchange interaction and it has intensively been studied in the literature.\cite{gammonprl,favero,flissikowski,bayerprb,karlsson} More precisely, it arises from the anisotropic part of the exchange interaction between an electron and a hole forming the exciton. The magnitude of FSS can be defined as $\delta_{\textrm {1}}^{\textrm X}\equiv |E_{\textrm {X}_{\parallel}}-E_{\textrm{X}_{\perp}}|$ \cite{bayerprb,favero,karlsson,flissikowski}. The splitting observed between two polarised components of the X emission line provides the value of $\delta_{\textrm {1}}^{\textrm X}$. The biexciton state is spin-degenerate, as a consequence of the spin-singlet state of the $e$-$h$ pairs it is composed of. The splitting of the 2X emission line is therefore related to the~FSS of the exciton. This in turn means that the splitting of the 2X and X emission lines have the same magnitude but different polarization axes which extends over mutually perpendicular directions (see \mbox{Fig. \ref{fig:anizomikropl}(a)} and \ref{fig:anizomikropl}(b)).\cite{akopian}  The~anisotropy axis can be defined as an angle of a polarization analyser at which the intensity of the low-energy component of the X emission line is maximized while its high-energy component disappears. This angle has been previously ascribed to the [110] crystallographic direction of the studied sample.\cite{molasnatural} 

The emission lines associated with the $p$-shell, the 3X and the 4X, are also split into two linearly polarized components, as it is shown in Fig.  \ref{fig:anizomikropl}(c) and \ref{fig:anizomikropl}(d). The magnitudes of the FSS measured for those lines are equal and their polarization axes are oriented along mutually perpendicular directions. Moreover, the respective polarization axes of the neutral exciton and triexciton as well as of the biexciton and quadexciton states are parallel. It means that the triexciton state has a non-zero FSS similarly to the neutral exciton state. On the other hand, the quadexciton state and the biexciton state do not exhibit the~FSS. In consequence, the FSS of the triexciton state can be defined as $\delta_{\textrm {1}}^{\textrm {3X}}\equiv |E_{\textrm {3X}_{\parallel}}-E_{\textrm {3X}_{\perp}}|$. Nevertheless, it must be taken into account that the fine structure of the triexciton state is related to the exchange interaction between the electron and the hole occupying the $p$-shell levels in a QD with an anisotropic form of the confining potential.

The theoretical description of the~triexciton state is more complex than for the~exciton state due to a difference between the $p$- and $s$-shell levels. In~QDs with symmetrical confining potential, there are two degenerate SP states with a $p$-like envelope function: the $p_+$ and the $p_-$, which refer to the angular momentum quantum number +1 and -1 respectively. This leads to the formation of four excitonic configurations of the triexciton which posses the lowest energy ground state.\cite{hawrylak} In close analogy to the exciton, the $e$-$h$ Coulomb interaction splits them into two configurations: an optically active (bright) and inactive (dark). The asymmetry of the confining potential affects the two degenerate bright states of the triexciton and splits them into two linearly polarized components, as it was predicted on a theoretical ground \cite{chithraniE} and experimentally demonstrated.\cite{molastri} In the case of quadexciton, there are two configurations regarding the spin of electrons and the angular momentum of holes occupying the $p$-shell levels: the~spin-singlet states (two electrons and two holes are anti-parallel) and the spin-triplet states (two electrons and two holes are parallel). \cite{hawrylak} In the~former case, the anisotropy of the confining potential increases the number of bright excitonic states from two to five, while in the latter one the basic set consists of one configuration only, in which there are two spin-parallel electrons and two angular-momentum-parallel holes distributed on two $p$-shell levels and there is no qualitative difference between the isotropic and anisotropic confinement.\cite{chithraniE} Due~to~large splitting between the $p$-subshells levels observed in the dots under investigation at zero magnetic field, the description of the quadexciton state becomes more complex than discussed above. In spite of that, based on the polarization pattern of the 4X emission line, it can be concluded that the~quadexciton state is not split in the studied case. 

\begin{figure}
\includegraphics[width=85mm]{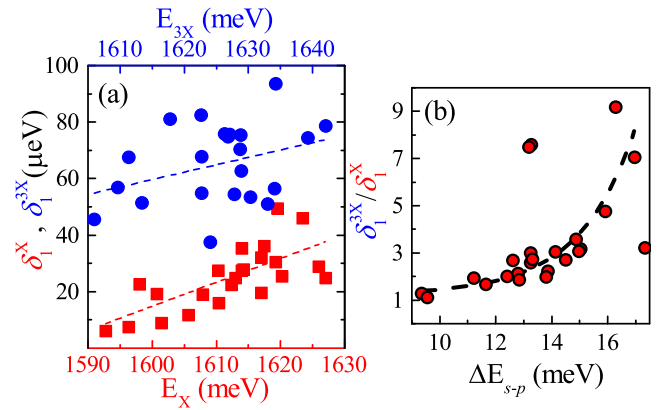}
\caption{(color online) (a) Fine structure splitting $\delta_{\textrm {1}}^{\textrm X}$ ($\delta_{\textrm {1}}^{\textrm {3X}}$) as a function of the X (3X) emission energy for a series of single QDs. (b) The ratio $\delta_{\textrm {1}}^{\textrm {3X}}/\delta_{\textrm {1}}^{\textrm X}$ as a function of the energy separation $\Delta \textrm{E}_{s-p}$. Dashed lines are guides to the~eye. \label{fig:delta}}
\end{figure}

In Fig. \ref{fig:delta}, we present the statistics of the amplitude of the $\delta_{\textrm {1}}^{\textrm X}$ and$\delta_{\textrm {1}}^{\textrm {3X}}$ splittings on our ensemble of ~20 single QDs. The parameter $\delta_{\textrm {1}}^{\textrm X}$, which describes the FSS of the exciton, varies from 5 \textmu eV up to 50 \textmu eV in the whole energy range. In the case of the triexciton, an overall change in $\delta_{\textrm {1}}^{\textrm {3X}}$ which covers the range from 35 \textmu eV to 95 \textmu eV is five times smaller. We may conclude some trend for both $\delta_{\textrm {1}}^{\textrm X}$ and $\delta_{\textrm {1}}^{\textrm {3X}}$ to increase with the characteristic emission energy of the dot - see Fig. \ref{fig:delta}(a), though this conjecture is not very well defined. Better pronounced correlation is found between the ratio of the $\delta_{\textrm {1}}^{\textrm X}$ to $\delta_{\textrm {1}}^{\textrm {3X}}$ splitting with the energy separation the $s$- and $p$-shell levels, defined as $\Delta_{s\textrm{-}p}=E_{\textrm {3X}}-E_{\textrm X}$. This ratio monotonically increases from $\sim$1 for $\Delta_{s\textrm{-}p}$=9 meV up to $\sim$9 for $\Delta_{s\textrm{-}p}$=17 meV.  In our opinion, this relationship can be understood in terms of different spatial extent of the $s$- and $p$-shell orbitals, but additional theoretical studies are necessary to support such a supposition.

\section{Conclusions \label{sec:summary}}

The micro-PL spectra of single GaAlAs/AlAs QDs have been measured. The emission lines attributed to the recombination of the quadexciton, triexciton, biexciton, and exciton state have been identified and analysed. Each emission line associated with the quadexciton cascade shows a doublet structure in polarization-resolved experiments. A pair of the 4X and 3X emission doublets is found to follow the same polarization rules as a pair of the 2X and X emission doublets. The amplitude of the doublet splitting is the same within each pair what points out the absence of the FSS not only in the biexciton but in the quadexciton state, as well. The FSS in the triexciton state surpasses the FSS in the exciton state by a factor up to eight and their ratio scales with the energy distance between the 3X and X emission lines, thus, very likely, with a lateral size and/or a composition of the dot. Based on these findings, the ladder of the excitonic states is proposed that explains the polarization properties of the emission lines observed in the experiment.

\begin{acknowledgments}
The work has been supported by the Foundation for Polish Science International PhD Projects Programme co-financed by the EU European Regional Development Fund and by EuroMagNETII (JRA8) under the EU contract 228043. MM kindly acknowledges the National Science Center (decisions DEC-2013/08/T/ST3/00665 and DEC-2013/09/N/ST3\linebreak[4]/04237) for financial support for his PhD.
\end{acknowledgments}

\bibliographystyle{apsrev4-1}
\bibliography{quabi}

\end{document}